\documentclass[conference]{IEEEtran}
\IEEEoverridecommandlockouts

\usepackage[utf8]{inputenc}
\usepackage{listings}
\usepackage{caption}
\usepackage{tabularx}
\usepackage{enumitem}
\usepackage{multirow}
\usepackage[dvips]{graphicx}
\usepackage[dvips]{color}
\usepackage[bookmarks=false,hidelinks]{hyperref}
\usepackage{todonotes}

\definecolor{dkgreen}{rgb}{0,0.6,0}
\definecolor{gray}{rgb}{0.5,0.5,0.5}
\definecolor{mauve}{rgb}{0.58,0,0.82}

\usepackage{tcolorbox}
\usepackage{booktabs}
\newcommand{\rb}[1]{

\vspace{0.3cm}
\begin{tcolorbox}[colback=gray!05,
                  colframe=black,
                  width=\columnwidth,
                  arc=3mm, auto outer arc,
                  boxrule=0.5pt,
                 ]
  #1
\end{tcolorbox}
}

\newcounter{Finding}
\stepcounter{Finding}

\newcommand{\implicationo}[1]{
	\rb{
	\noindent
	\textit{\textbf{Finding \theFinding}. #1}
	}
	\stepcounter{Finding}
}

\begin{document}
\begin{sloppy}

\title{Comprehending Test Code: An Empirical Study}

\author{\IEEEauthorblockN{Chak Shun Yu}
\IEEEauthorblockA{\textit{Department of Software Technology} \\
\textit{Delft University of Technology}\\
Delft, the Netherlands \\
chakshunyu@gmail.com}
\and
\IEEEauthorblockN{Christoph Treude}
\IEEEauthorblockA{\textit{School of Computer Science} \\
\textit{University of Adelaide}\\
Adelaide, Australia \\
christoph.treude@adelaide.edu.au}
\and
\IEEEauthorblockN{Maur\'{i}cio Aniche}
\IEEEauthorblockA{\textit{Department of Software Technology} \\
\textit{Delft University of Technology}\\
Delft, the Netherlands \\
M.FinavaroAniche@tudelft.nl}
}

\maketitle

\newcommand{\jpacman}{JPacman}
\newcommand{\pacman}{Pacman}

\begin{abstract}

Developers spend a large portion of their time and effort on comprehending source code. While many studies have investigated how developers approach these comprehension tasks and what factors influence their success, less is known about how developers comprehend test code specifically, despite the undisputed importance of testing. In this paper, we report on the results of an empirical study with 44 developers to understand which factors influence developers when comprehending Java test code. We measured three dependent variables: the total time spent reading a test suite, the ability to identify the overall purpose of a test suite, and the ability to produce additional test cases to extend a test suite. The main findings of our study, with several implications for future research and practitioners, are that (i) prior knowledge of the software project decreases the total reading time, (ii) experience with Java affects the proportion of time spent on the Arrange and Assert sections of test cases, (iii) experience with Java and prior knowledge of the software project positively influence the ability to produce additional test cases of certain categories, and (iv) experience with automated tests is an influential factor towards understanding and extending an automated test suite.

\end{abstract}


\section{\label{cha:intro}Introduction and Motivation}

An essential aspect of software development is being able to understand how a program works~\cite{siegmund2016program,Schroter:2017:CSP:3101414.3101451}, resulting in developers spending the most significant portion of their time on reading and understanding source code~\cite{siegmund2016program, crosby2002roles, Maalej:2014:CPC:2668018.2622669}. The software engineering research community has investigated the process of program comprehension from various angles. While no single general approach exists to explain the process of program comprehension of developers in its entirety~\cite{Maalej:2014:CPC:2668018.2622669}, studies have investigated the possible influences of various aspects, ranging from how to approach and improve program comprehension from a holistic point of view~\cite{siegmund2016program, Schroter:2017:CSP:3101414.3101451, Maalej:2014:CPC:2668018.2622669} to properties of a system at source code level (e.g., style, quality, and length of identifier names~\cite{Hofmeister2018, hofmeister2017comparing, Sharif:camelcase, Fakhoury}), at code construct level (e.g., code regularity~\cite{Jbara:2014:ECR:2597008.2597140} and code beacons~\cite{crosby2002roles, kelly2015using}), and at higher levels of abstraction (e.g., code smells~\cite{Soh:7476660}, readability~\cite{Posnett:2011:SMS:1985441.1985454}, and familiarity~\cite{kruger2018you}). Others have investigated the role of code visualizations~\cite{bachercode, Mumtazsmells, Cornelissen:2007:VTA:1251979.1252787, Kienle4721178} and drawn parallels with natural language comprehension~\cite{Busjahn:2015:EMC:2820282.2820320, Peitek:2018:SMP:3239235.3240495, Peitek8425769Heads}.

These studies provide important contributions towards our knowledge on program comprehension and together form our current understanding of program comprehension. However, less is known about how developers comprehend test code and which factors are of influence on this comprehension, despite the undisputed importance of tests, e.g., to improve the quality of software projects, to ensure correct behaviour, or as documentation~\cite{Cornelissen:2007:VTA:1251979.1252787, spadini2018relation, Greiler:2012:MTC:2366988.2366996, fucci2018longitudinal, Fucci7592412, Zhu:1997:SUT:267580.267590}. Developers often disregard test code during software maintenance activities because it tends to be complex and costly~\cite{spadini2018relation, Greiler:2012:MTC:2366988.2366996, Spadini:256e7d56352f44ae919b97fad0eafe69}, causing the quality and usefulness of test code to decrease over time~\cite{spadini2018relation, DeursenMBK01, Moonen2008, Bavota:2015:TSR:2790618.2790636}.

To combat this, several studies have investigated ways of enhancing the test maintenance process for developers~\cite{Cornelissen:2007:VTA:1251979.1252787, spadini2018relation, Greiler:2012:MTC:2366988.2366996, DeursenMBK01, Bavota:2015:TSR:2790618.2790636}. While these studies present ways to enhance the process of test code comprehension and while the theories behind their enhancements are valuable contributions towards understanding test code comprehension, none of them set their primary focus on learning more about the underlying process of test code comprehension. However, it is necessary to gain a better understanding of how developers approach the comprehension of tests to be able to improve it.

In this work, we fill this gap by investigating the factors which influence developers' test code comprehension. We conducted an exploratory empirical study with 44 participants and a total of 132 data points, defining three metrics represented across nine dependent variables: the amount of time spent on reading a test suite, the ability to identify the overall purpose of a test suite, and the ability to extend a test suite by producing additional test cases of varying categories.

The main findings of our study are that (i) prior knowledge of the software project decreases the total reading time, (ii) experience with the programming language affects the proportion of time spent on the Arrange and Assert sections of test cases, (iii) experience with the programming language and prior knowledge of the software project positively influence the ability to produce additional test cases of certain categories, and (iv) experience with automated tests is an influential factor towards understanding and extending an automated test suite.

The contributions of this paper are:

\begin{itemize}

\item An exploratory empirical study to understand the factors influencing test code comprehension. The study reveals a collection of factors that influence test code comprehension along with their impact.

\item A set of quantifiable metrics to measure test code comprehension that can be used in future studies.

\end{itemize}

\section{\label{cha:related}Related Work}

Test code comprehension is not yet well explored in software engineering, contrary to the more general domain of program comprehension. This section discusses related studies from the domain of program comprehension, and the few existing studies related to test code comprehension. We summarize and group them by their research methods in Table~\ref{tbl:works_overview}.

\begin{table}
\centering
\caption{Overview of related studies.}
\label{tbl:works_overview}
\begin{tabular}{llp{6.4cm}}
\toprule
 & Work & Contribution \\ \midrule
\parbox[t]{2mm}{\multirow{10}{*}{\rotatebox[origin=c]{90}{Conventional}}} 
    & \cite{bachercode}  
        & Enhancing mini-map visualization with layer of scope chain information. \\
    & \cite{Bavota:2015:TSR:2790618.2790636}  
        & Assessing whether developers are able to identify test smells (initial survey). \\
    & \cite{Cornelissen:2007:VTA:1251979.1252787}  
        & Use scenario diagrams to assist test code comprehension. \\
    & \cite{kruger2018you}  
        & Impact of code familiarity on software development. \\
    & \cite{Maalej:2014:CPC:2668018.2622669}  
        & Study on how developers approach program comprehension. \\
    & \cite{Mumtazsmells}  
        & Use multivariate data visualization techniques to identify bad smells. \\  \midrule
\parbox[t]{2mm}{\multirow{10}{*}{\rotatebox[origin=c]{90}{Eye Tracking}}} 
    & \cite{Busjahn:2015:EMC:2820282.2820320} 
        & Linearity of developers' source code reading behaviour. \\
    & \cite{crosby2002roles} 
        & Role of experience during program comprehension. \\ 
    & \cite{kelly2015using} 
        & Using knowledge of experts to assist novice developers. \\ 
    & \cite{hofmeister2017comparing} 
        & Comparing eye movements between experts and novices during program comprehension. \\ 
    & \cite{Jbara:2014:ECR:2597008.2597140, Jbara:2015:PRR:2820282.2820319} 
        & Impact of code regularity on effort and complexity. \\
    & \cite{Sharif:camelcase} 
        & Impact of identifier naming conventions on program comprehension. \\ 
    & \cite{Melo:2017:VTE:3101414.3101419} 
        & Impact of variability on debugging. \\
    & \cite{Sharif:2012:ESR:2168556.2168642} 
        & Impact of (initial) scan time on finding code defects. \\  \midrule
\parbox[t]{2mm}{\multirow{5}{*}{\rotatebox[origin=c]{90}{Cognitive}}}      
    & \cite{Fakhoury} 
        & Impact of poor code lexicon on cognitive effort. \\
    & \cite{Fritz:2014:UPM:2568225.2568266} 
        & Detecting when tasks in software development are considered difficult by developers. \\
    & \cite{KOSTI201852} 
        & Cognitive workload during program comprehension to asses task difficulty. \\  \midrule
\parbox[t]{2mm}{\multirow{5}{*}{\rotatebox[origin=c]{90}{Others}}}       
    & \cite{Hofmeister2018} 
        & Impact of identifier name length on program comprehension. \\ 
    & \cite{Schankin:2018:DCI:3196321.3196332} 
        & Comparing the impact of compound and short identifier names on program comprehension.  \\ 
    & \cite{Soh:7476660}
        & Impact of code smells on program comprehension. \\ 
\bottomrule
\end{tabular}
\end{table}

\subsection{Impact of Code Lexicon on Program Comprehension}

In their studies, Hofmeister et al.~\cite{Hofmeister2018} and Schankin et al.~\cite{Schankin:2018:DCI:3196321.3196332} investigated the impact of the length of identifier names on the program comprehension of developers. The results of their study indicate that shortening identifier names to abbreviations negatively impacts program comprehension. Work by Fakhoury et al.~\cite{Fakhoury} uses an alternative form of fMRI---functional near infrared spectroscopy (fNIRS)---in combination with eye tracking to measure the effects of a poor source code lexicon on the cognitive effort required by developers in the process of program comprehension. Results indicate that a poor source code lexicon, in any form, has a negative impact on the comprehension of the respective source code and the ability of developers to perform their software development tasks. In another study, Kosti et al.~\cite{KOSTI201852} worked towards a way to asses the cognitive workload of developers during software development tasks and a model to assess the difficulty of these tasks. To do so, they employed the research method of Electroencephalography (EEG). Comparisons between the EEG patterns of all participants indicated clear differences in the cognitive workload during code comprehension and finding syntax errors.

\subsection{Code Constructs and Program Comprehension}

Logically, more complex source code is more difficult to comprehend than simpler code. While metrics exist to measure the complexity of source code, such as lines of code (LOC) and McCabe's cyclomatic complexity, studies by Jbara and Feitelson~\cite{Jbara:2014:ECR:2597008.2597140, Jbara:2015:PRR:2820282.2820319} identified a mismatch between the theory behind these metrics and how complexity is interpreted in practice in certain cases. In their work, the authors introduce the notion of code regularity---a repetitive code segment with potential small adjustments in every iteration---and measure the potential effects it has on code complexity and thus code comprehension. To do so, eye tracking technology was used to determine the differences in effort used by developers to comprehend code snippets with varying levels of regularity, which was measured by the time and number of fixations spent. Results indicated that a high rate of regularity in code snippets has no impact on the time spent on the tasks, but does lead to better task performance and comprehension. Moreover, the authors found a diminishing amount of effort with every repetition of a code block, based on which they concluded that the additive nature of the syntactic complexity metrics causes an overestimation of the complexity of regular code, and should be modified with context-dependent weights. 

Melo et al.~\cite{Melo:2017:VTE:3101414.3101419} used eye tracking to understand how developers debug programs with variability, which in short is the presence of configuration-dependencies at compile time. The authors found that variability increases the debugging time for code fragments containing variability as well as code fragments in the proximity of variability-containing code fragments. The number of saccades between definition-usages of fields and call-returns for methods prolongs the initial scan of the program and splits the debugging approach of developers into either consecutive or simultaneous processing of the configurations. 

Work by Crosby et al.~\cite{crosby2002roles} investigated the impact of experience on program comprehension by looking at how different groups identify beacons (important code segments). The results indicate that developers identify beacons differently based on their experience. Experienced developers will focus more on identifying beacons in a software program, while novice developers are less likely to search for beacons. Building on this concept of beacons, work by Hegarty-Kelly et al.~\cite{kelly2015using} showed how knowledge on differences in performance (in identifying beacons) can be used to improve the process of program comprehension for certain groups of developers.

\subsection{Similarities with Natural Language Comprehension}

Busjahn et al.~\cite{Busjahn:2015:EMC:2820282.2820320} conducted an eye-tracking study on the linearity of developers' source code reading behaviour. The comparison between reading natural language text and source code has often been made. However, the authors identify that the linearity aspect, which is a significant property of natural language text, is left quite unexplored. In this study, Busjahn et al.~\cite{Busjahn:2015:EMC:2820282.2820320} made an attempt in exploring this aspect of linearity in source code reading behaviour. The results of their study show that novices read source code less linearly than natural language text---70\% linear eye movements compared to 80\%---and, on top of that, experts read source code less linearly than novices.

\subsection{Test Code Comprehension}

While program comprehension has been extensively studied, the comprehension of test code has not received the same amount of attention. Despite this, some studies have looked into possible ways to enhance the test code comprehension process of developers~\cite{Cornelissen:2007:VTA:1251979.1252787, Greiler:2012:MTC:2366988.2366996, Bavota:2015:TSR:2790618.2790636}.

Greiler et al.~\cite{Greiler:2012:MTC:2366988.2366996} investigated a way to derive relations between levels of test cases. Their approach attempts to connect higher level end-to-end tests to low level unit tests through similarity of their stack traces. This aids developers when changes occur in requirements, by making it easier for them to trace the changes from the end-to-end tests to the affected source code through unit tests.

Bavota et al.~\cite{Bavota:2015:TSR:2790618.2790636} investigated the origin and survivability of test smells and the relationships of their presence with production code smells. Based on the analysis of the commit history of 152 open source projects, their results indicate that test smells originate during the creation of the test cases rather than over time, that they have a high survivability, and that they have certain relationships with the presence of code smells.

Finally, work by Cornelissen et al.~\cite{Cornelissen:2007:VTA:1251979.1252787} introduced a visual approach to assisting developers' test code comprehension processes. Their approach creates scenario diagram models for test cases based on dynamic analysis of the test suites. These scenario diagrams focus on the interactions between objects, abstracting away unnecessary or less important information, and visualizing them in a human readable way. Based on their case study, they conclude that test code visualization in the form of scenario diagrams yields useful information regarding the system's inner workings.

\section{\label{cha:researchdesign}Research Design}

In this section, we outline our research questions and the overall methodology of our empirical study along with independent and dependent variables. We also describe our procedures for participant selection and data analysis.

\subsection{\label{sec:researchquestions}Research Questions}

The focus of this study lies in identifying and establishing potential relationships between (software development related) properties of developers and the degree to which they are able to comprehend source code tests. We associate three factors with test code comprehension: the developers' Reading Time (RT), their ability to Identify the Testing Purpose (ITP), and their ability to Produce Additional Cases (PAC).

\begin{description}

\item[RQ1] {\label{fac:time} \textit{What factors influence the time that developers spend reading test code?}} Reading time is the amount of time that developers spend on reading test code before moving on to the next task.

\item[RQ2] {\label{fac:purpose} \textit{What factors influence the ability of developers to identify the purpose of a test suite?}} An important aspect when reading test code is to understand the higher-level themes that are used to group test cases which exercise similar scenarios into test suites.

\item[RQ3] {\label{fac:cases} \textit{What factors influence the ability of developers to produce additional test cases to extend a test suite?}} In practice, understanding test code and its underlying purpose is not enough. After understanding the test code, it is often necessary to produce additional test cases.

\end{description}

\subsection{\label{sec:methodology}Methodology}

To answer our research questions, we conducted an empirical study where participants were invited to read test suites and answer questions about them. In the remainder of this section, we explain the decisions and trade-offs that we considered during the design of the study as well as the overall procedure that participants went through.

The entire study procedure as well as resulting data are publicly available in our online appendix~\cite{appendix}.

\subsubsection{\label{sec:pre_interview}Pre-Study Questionnaire}

Participants were asked to complete a pre-study questionnaire which asked for their gender, age, software development role, amount of experience with software development in years, amount of experience with Java in years, current programming language of choice, and amount of experience with automated test code (in this case, experience with JUnit tests). Their experience with Java and current programming language of choice together form a proxy for their familiarity and comfort with Java, which is important as the main programming language used in the study was Java.

\subsubsection{\label{sec:study}\label{sec:tool}Trials}

Participants were then asked to work on three trials, each with a different test code snippet, one at a time. We showed the specific instructions as well as a small tutorial on how to use our tool (see below) before starting the trials.

Each trial consisted of two parts. First, participants were given a test code snippet and were asked to read and comprehend it. We measured the time they spent on this task (to answer RQ1). Whenever participants felt that they had comprehended the test code, they moved to the next part of the trial, which consisted of questions about the tests they just read. Note that participants had no access to the test code anymore as soon as they saw the questions, to ensure they actually attempted to comprehend the code during the reading period. We asked two questions (corresponding to RQs~2 and~3):

\begin{enumerate}

\item What is the purpose of the test suite?

\item Describe additional test cases that you would write to extend the current test suite. Use one line per case.

\end{enumerate}

Besides measuring the time a participant spent reading the test code, we collected the time they spent in its different parts. Test code can be divided into three parts (also known as AAA~\cite{grig2012AAA}): \textit{Arrange}, where the inputs that will be passed to the method under test are decided; \textit{Act}, where the method under test is invoked; and \textit{Assert}, where the code verifies that the method behaved as expected. To collect such fine-grained data about what participants are looking at, our tool only allows participants to see five lines of code at a time. To see other lines, participants use their keyboard's arrow keys. Our tool collected the number of seconds each line was visible to a participant. 

In terms of functionality, the tool is similar to the design by Hofmeister et al.~\cite{Hofmeister2018}, however, there are differences in terms of configuration: While Hofmeister et al.~\cite{Hofmeister2018} configured the size of the viewport as approximately one third of the code snippet---seven lines of code---we set this number to five to ensure that the viewport does not cover too many different AAA parts, thus improving the accuracy of our data collection. We tested different numbers based on the specific test code snippets used in the study.

The order in which the three trials are provided to each participant is randomized to mitigate possible learning bias. To mitigate bias possibly resulting from developers' linear reading patterns---past work~\cite{Busjahn:2015:EMC:2820282.2820320} has shown that developers follow linear reading patterns when reading source code, although less than when reading natural language text---the starting viewport of every trial (i.e., the lines the participant could see when the trial started) is randomized across the test suite. We limited each trial to 10 minutes.

\subsubsection{\label{sec:tests}Test Code Selection Criteria}

We used test code from \jpacman, a software used to teach software testing in a Computer Science programme, in our study to fulfill the following selection criteria: 

\begin{enumerate}

\item The test classes are understandable in an isolated and standalone manner, which means that participants do not have to see the entire test suite (that can be composed of dozens of tests) to have a good grasp of its goals.

\item The test classes adhere to the AAA testing structure in an unambiguous manner, which reduces the cognitive load to understand the test classes, and enables us to measure the time spent in each of the AAA parts.

\item Most people already know the concept of \pacman, which reduces the cognitive effort required from participants to understand the underlying domain.

\end{enumerate}

To select test code snippets from the test code of \jpacman, we computed the number of lines of code of all test methods in the project, and selected one test case from the first quartile (six lines), one from the median (nine lines), and one from the third quartile (fourteen lines). We selected test cases which covered the three parts of the AAA pattern evenly, to ensure that lines from each part had a similar chance of being read.

To support participants in identifying the purpose of a test suite, we included one additional test case from the same suite with each test case (i.e., participants had access to two test cases per test suite). We ensured that test cases were self-contained (e.g., not relying on variables declared outside of the method) and anonymized the names of the test cases to mitigate possible bias stemming from the clarity or lack of clarity of these names. The final test snippets can be found in our appendix.

\subsection{\label{sec:independent-variables}Independent Variables}

The independent variables in our research are mainly static properties which we expected to be relevant based on previous work. We discuss our expectations regarding their potential influence on the different metrics of test code comprehension below.

\subsubsection{Participant Age} 

While age has been a common subject of research in the field of linguistics regarding its effect on language comprehension~\cite{cenoz2003influence, LIDZBA20116}, it has rarely been studied in the field of Computer Science. Our expectations are that the age of participants will not have any significant effects on their test code comprehension.

\subsubsection{Participant Gender}

Gender has already been subject of research in the field of Computer Science in previous studies~\cite{6240505, Beckwith:2005:EED:1054972.1055094, Burnett:2010:GDP:1852786.1852824, lawrie2007effective}. Researchers have found that the main difference between individuals of different genders lies in the way in which software development related tasks are approached, while the results and performance show no correlation with gender. We expect that, similar to previous studies, gender will have no influence on performance in our study. However, we expect that differences might be observed with regard to the amount of time that participants spend on reading tests. A study by Sharafi et al.~\cite{6240505} showed that female subjects take more time to carefully elaborate on their decisions.

\subsubsection{Participant Experience}

Many researchers have investigated the relationship between program comprehension and developers' experience level~\cite{crosby2002roles, hofmeister2017comparing, Sharif:camelcase, kelly2015using, Busjahn:2015:EMC:2820282.2820320, peachock2017investigating, kontogiorgos2015towards, pavel2015experts, tvarozek2015studying}. In this research, we consider three types of experience, with the expectation that similar to the findings of previous work, more experience will positively affect the task performance and the degree of code comprehension~\cite{crosby2002roles, Jbara:2014:ECR:2597008.2597140, kelly2015using, Jbara:2015:PRR:2820282.2820319}: 

\begin{itemize}

\item experience as a \textit{developer}, 

\item experience with the \textit{Java} programming language, and 

\item experience in using automated \textit{test code}. 

\end{itemize}

A previous study by Peitek et al.~\cite{Peitek:2018:SMP:3239235.3240495} has already highlighted the positive influence of familiarity with the programming language on program comprehension.

\subsubsection{Prior Knowledge of Software Project}

This binary variable indicates whether a participant has prior knowledge on the specific software project. We expect that project knowledge will affect the dependent variables positively.

\subsubsection{Participant UUID and Trial number}

The unique identifier for the participant and the trial number. These two variables are used as random factors in our models (see Section~\ref{sec:analysis_procedure}).

\subsection{\label{sec:dependent-variables}Dependent Variables}

In this section, we introduce the dependent variables we use in our models.

\subsubsection{TotalTimeInSecs}

The amount of time the participant spent reading the provided test code snippet, in seconds.

\subsubsection{\%Arrange, \%Act, \%Assert}

The percentage of time the participant spent in each of the three AAA parts of the test code snippet.

\subsubsection{Identification of the Testing Purpose (ITP)}

A binary variable that captures whether the participant has correctly identified the purpose of the provided test suite. We manually analyzed participants' answers to question 1 (see Section~\ref{sec:study}) and systematically assigned a 0 (not identified) or a 1 (identified). This work was done by the first author of this paper, who is an expert on \jpacman's source code. 

As examples, participants gave us answers such as \textit{``I believe the purpose of this test suite is to minimally check whether the factory objects are capable of creating a level and a ghost.''} and \textit{``to check that when a player eats the last pellet, the player wins and if a player eats a ghost, the game will end and the player is dead''}.

\subsubsection{Producing Additional Cases (PAC)}

A set of binary variables that indicate whether, and in which way, participants were able to extend the test suite. We analyzed the data in two steps. First, we analyzed all participants' answers in order to categorize the type of tests they were able to create. For each answer, we either decided that an already emerged category would fit, or we created a new category. The categories were then refined together with the third author of this paper (who is also an expert on \jpacman's source code and has more than ten years of experience in automated software testing).

Participants' answers were written in natural language (i.e., not source code). Examples of answers included \textit{``create levels of different sizes''} and \textit{``assert that ghosts are initialized on their desired spawn location''}. Given that the field was required, we also received answers such as \textit{``I can't think of any additional test cases for this test suite''}, indicating that a participant was not able to devise more test cases (which does not fit into any of the following categories).

The following four categories emerged from this analysis:

\begin{itemize}

\item \textit{Basic tests.} Extension of the provided cases in a limited manner based on the information in the provided cases.

\item \textit{Domain tests.} Extending cases to test valid scenarios and input, but not limited to examples given in the provided cases.

\item \textit{Error tests.} Additional cases with invalid input or dedicated to fail.

\item \textit{Other tests.} Tests that did not fit any of the previous categories.

\end{itemize}

As a second step, we then assigned values to four binary variables (0s and 1s), related to the four categories above. For each participant $P$ and category $C$, we assigned a 1 if $P$ was able to produce one or more tests for category $C$; otherwise, we assigned 0.

\subsection{\label{sec:participants}Participant Selection}

For this research, we recruited two groups of participants: a group that had prior knowledge of the software project (i.e., knows the source code of the project) and another group that did not have this prior knowledge. We recruited participants belonging to the former group in a university course which uses the software project from which we draw test cases for the study. These students had two months of experience with the project's code base. Participation was voluntarily, and did not affect students' grades. For the latter group, the study was shared on social media platforms to invite developers to participate. We did not have any further selection criteria, aiming to recruit a diverse set of participants.

\subsection{\label{sec:analysis_procedure}Analysis Procedure}

We introduce our data analysis methodology in this section.

\paragraph{Reading Time}

To investigate the influence of the independent variables on reading time (RT), we used Linear Mixed Model (LMM) analysis. For each dependent variable (\textit{totalTimeInSecs, \%Arrange, \%Act}, and \textit{\%Assert}), an LMM was constructed with all independent variables as fixed effects. As each participant completed three trials in our study and these three trials were the same for every participant, patterns caused by this overlap in trials and participants can affect the results of our models. To mitigate this, both variables (\textit{trial number} and \textit{participant uuid}) were represented as random effects in the models.

The resulting model for an LMM analysis relies on several assumptions: \textit{(i)} linearity in the data of the model, \textit{(ii)} there should be no collinearity between fixed effects, \textit{(iii)} absence of heteroskedasticity, which means that the residuals in the model need to have a similar amount of variation for all predicted values, \textit{(iv)} the residuals of the LMM model need to be normally distributed, \textit{(v)} there should be no influential data points, and \textit{(vi)} independence should hold across the data of the model. 

We verified Assumptions \textit{(i), (iii),} and \textit{(iv)} by inspecting visual plots of the residuals, and we verified Assumption \textit{(ii)} using visual plots of the linearity between every pair of independent variables. Assumption \textit{(v)} was verified through manual inspection and comparing the full model against reduced models, which revealed no significant differences. Finally, Assumption \textit{(vi)} was adhered to by conforming to a mixed effect model, rather than just a linear model. 

All models were created with R~\cite{r} and the \textit{lme4} package~\cite{lme4}. We verified the fitness of the models using marginal and conditional $R^2$ values~\cite{r2}, using implementations of the \textit{MuMIn} package~\cite{MuMIn}. For every independent variable, we conducted a likelihood ratio test of the full model against a reduced model without the effect in question. As is common with testing statistical significance, independent variables were deemed influential over the dependent variable when the probability of committing a Type-I error was at most 5\% ($\alpha = 0.05$).

\paragraph{Identifying Testing Purpose (ITP) and Producing Additional Cases (PAC)}

Contrary to the RT, the variables capturing the participants' abilities to identify the testing purpose (\textit{purposeScore}), and to produce additional cases (\textit{BasicTest, DomainTest, ErrorTest, and OtherTest}) are binary variables. Therefore, we performed a Binomial Logistic Regression (BLR) analysis to investigate the influence of the independent variables on ITP and PAC.

We modeled the five dependent variables (\textit{purposeScore, BasicCase, DomainCase, ErrorCase}, and \textit{OtherCase}) with the same collection of fixed effects, i.e., the previously described independent variables. The variables \textit{RT, \%Arrange, \%Act,} and \textit{\%Assert} are also included in the models as fixed effects. To verify and assess the resulting models, we use McFadden's pseudo-$R^2$ and analyze the deviance tables. Similar to RT, independent variables are deemed statistically significant and thus influential on the dependent variables when $\alpha \leq 0.05$.

\section{\label{cha:results}Results}

In this section, we report on the results of the empirical study.

\subsection{\label{sec:stats:participants}Participant Statistics}

After three months of hosting the online study, a total of 44 developers participated in our research, 10 (23\%) of which had prior knowledge of the code base. Furthermore, 9 (20\%) of the participants were female and roughly 86\% of the participants were either a developer (39\%) or a student (48\%). By far the most preferred programming language by the participants was Java (43\%), followed by Python (11\%) and C\# (9\%).

Table~\ref{tbl:participants_stats} shows descriptive statistics of the participants, separately for those with prior knowledge of the project and for those without. Participants with prior knowledge of the software project generally have less experience, which can be expected based on the inclusion criteria of the group with project knowledge (i.e., Computer Science students).

\begin{table*}
\centering
\caption{Descriptive statistics of participants, in years of experience.}
\label{tbl:participants_stats}
\begin{tabular}{l|rrrrr|rrrrr}
\toprule
 & \multicolumn{5}{c|}{Knowledge of the software project} & \multicolumn{5}{c}{No knowledge of the software project} \\ 
Factor & Min & Median & Mean & SD & Max & Min & Median & Mean & SD & Max \\ \midrule
Age & 18 & 19 & 19 & 1.44 & 23 & 19 & 24 & 26.7 & 5.94 & 43 \\
Developer & 1.0 & 2.5 & 3.2 & 2.12 & 7.0 & 0.0 & 5.5 & 7.37 & 5.40 & 20 \\
Java & 1.0 & 2.5 & 3.0 & 2.18 & 7.0 & 0.0 & 4.0 & 4.59 & 3.90 & 15 \\
Tests & 0.0 & 1.0 & 1.1 & 0.84 & 3.0 & 0.0 & 4.0 & 4.24 & 3.83 & 18 \\ \bottomrule
\end{tabular}
\end{table*}

\subsection{\label{sec:stats:quantification}Descriptive Statistics of the Dependent Variables}

For the ITP variables, the results in Table~\ref{tbl:quantification_stats_itp} show that every trial (i.e., the different test snippets) had a similar success rate. Roughly 18\% to 32\% of the participants correctly identified the general testing purpose of the respective test suite.

Table~\ref{tbl:quantification_stats_pac} shows the descriptive statistics for PAC. Participants are evenly capable of producing Basic (55\% of participants produced at least one basic case) and Domain (47\%) test cases, while they are less likely to produce Error (9\%) and Other (19\%) test cases.

\begin{table}
\centering
\caption{Distribution of the dependent binary variables for the ability of participants to identify testing purpose (ITP). Success means that the participant was able to identify the purpose of that test suite. Participants = 44.}
\label{tbl:quantification_stats_itp}
\begin{tabular}{lrrr}
\toprule
 & Trial 1 & Trial 2 & Trial 3  \\ 
 & (small snippet) & (medium snippet) & (long snippet) \\
 \midrule
Success & 8 & 14 & 8  \\
Fail & 36 & 30 & 36  \\
\bottomrule
\end{tabular}
\end{table}

\begin{table}
\centering
\caption{Distribution of the dependent binary variables for the ability of participants to produce additional cases (PAC). Success means that a participant was able to derive at least one test case for that category. Numbers in each category add up to 132 (i.e., 3 trials per participant, 44 participants).}
\label{tbl:quantification_stats_pac}
\begin{tabular}{lrrrr}
\toprule
 & Basic tests & Domain tests & Error tests & Other tests \\ 
 \midrule
Success  & 73 & 62 & 12 & 25\\  
Fail & 59 & 70 & 120 & 107\\
\bottomrule
\end{tabular}
\end{table}

Table~\ref{tbl:dependent_numerical} shows the distribution statistics of all numerical dependent variables, i.e., the proportions of time that participants spent on each AAA section during the study and the total amount of reading time spent on the test suite. The range for reading time is wide, from the minimum of less than half a minute (25.74 seconds) to the maximum of almost 10 minutes (590.35 seconds).

We can also observe that participants generally spent the least amount of time on the Act section. The Arrange section of tests was where most participants spent most of their time, compared to the other AAA sections, but this section also had the largest absolute differences between participants. While the proportion of time that participants spent on the Assert section is generally in between the other two sections, there are outliers in either direction, leading to more time than the Arrange section or less time than the Act section.

\begin{table}
\centering
\caption{Distribution of the numerical dependent variables.}
\label{tbl:dependent_numerical}
\begin{tabular}{lrrrrr}
\toprule
Factor & Min & Median & Mean & SD & Max \\ 
\midrule
\%Arrange    & 15.27        & 48.17           & 49.20         & 16.94       & 82.70        \\
\%Act        & 7.38         & 20.11           & 20.45         & 5.96        & 42.10        \\
\%Assert     & 4.26         & 34.12           & 30.35         & 17.31       & 61.70        \\
Reading Time (sec)        & 25.74        & 93.04           & 129.59        & 107.28      & 590.35       \\ 
\bottomrule
\end{tabular}
\end{table}

\subsection{\label{sec:verification_stats}Model Assumptions}

We observed that there is a negative correlation between the proportion of time that participants spent on the Arrange section and the Assert section of a test (data in the appendix). Since none of these two variables shows any significant difference in the scatterplots with the other independent variables, there are no clear benefits of choosing either of them. Without any particular deterministic reasoning, the proportion of time that participants spent on the Assert section was chosen over the Arrange section for the RT, ITP, and PAC models.

Manual inspection of the histogram and the Q-Q plot of the reading time LMM (in the appendix) indicates violations to the linearity of the model and the required normal distribution of residuals. When inspecting the residuals against the fitted values of the model however, there is a noticeable pattern in the graph: Higher fitted values have larger residuals, indicating that the variance is larger in the higher range and smaller in the lower range. This violates the assumed absence of heteroskedasticity and thus renders the model inaccurate.

To address this violation, we applied one of the most common solutions by conducting a log transformation of the RT variable. Statistical and graphical analysis showed that the RT data is log-normally distributed, justifying this choice. This is further supported by re-inspecting the histogram and the Q-Q plot of the newly created model (in the appendix), which are improved compared to the original model and display better indications of linearity of the model and normality of the residuals. Furthermore, the residual plot against the fitted values of the new model shows no marginal pattern in the variations of the residuals.

\subsection{\label{sec:model_stats}Analysis of the Final Model}

\begin{table*}
\centering
\caption{Models represented as rows with independent variables as columns. The values are the coefficients of the fixed effects on the dependent variable. Asterisks indicate statistically significant effects ($\alpha \leq 0.05$).}
\label{tbl:results_models}
\begin{tabular}{lrrrrrrrrr}
\toprule
Predicted & & & & & & Prior & & & \\
Variables   & Age & Gender & Develop & Java & Tests & Knowledge & \%Act & \%Assert & RT \\ 
\midrule 
RT (log)           & 0.005        & -0.090          & -0.002              & -0.010         & -0.013          & \textbf{-0.259}*          & -0.002         & 0.002         & --           \\ 
\%Arrange          & -0.442       & -4.300          & -0.092              & \textbf{0.867}*        & 0.051          & -0.480           & --             & --             & 0.008       \\ 
\%Act              & 0.056        & 1.930            & 0.169              & -0.223         & -0.121          & -2.285           & --             & --             & -0.001       \\ 
\%Assert           & 0.383        & 2.408           & -0.073              & \textbf{-0.636}*        & 0.071          & 2.895           & --             & --             & -0.005       \\ \midrule
purposeScore (ITP)       & 0.063        & 0.018           & 0.007              & -0.135         & \textbf{0.272}*         & 0.773           & 0.046         & -0.038         & -0.000       \\ 
BasicTest          & -0.016        & 0.042           & -0.063              & 0.106         & 0.096          & 0.194           & -0.067         & 0.021         & 0.000       \\ 
DomainTest         & 0.021        & 0.077           & -0.090              & -0.051         & \textbf{0.339}*         & \textbf{1.975}*          & -0.001         & -0.037         & 0.000       \\ 
ErrorTest          & -0.168        & 0.285           & -0.061              & \textbf{0.345}*        & 0.236          & 1.982           & 0.159         & -0.019         & \textbf{1.028e-5}* \\ 
OtherTest       & -0.129        & 1.224           & -0.030              & 0.049         & \textbf{0.266}*         & 0.478           & 0.021         & 0.008         & -0.000       \\ 
\bottomrule
\end{tabular}
\end{table*}

The results of the LMM and BLR models are reported in Table~\ref{tbl:results_models} in the form of p-values of all fixed effects per dependent variable.

The demographic variables---age and gender of the participants---have no statistically significant influence on any of the predicted variables. The experience of the participant as a developer and the proportion of time that they spent on each AAA section also have no statistically significant effect on any of the dependent variables. Table~\ref{tbl:significant_variables} shows all relevant data for statistically significant effects. In the following, we discuss each model in detail.

\subsubsection{Reading Time (RT)}

Our results do not indicate statistically significant effects of the different kinds of experience (as a developer, with Java, and with using tests) on the dependent variable of reading time (RT). The only statistically significant independent variable (albeit with a small effect) is whether the participant has prior knowledge of the software project that the test suites are targeted at. The log transformed RT is negatively affected by prior knowledge. 

\subsubsection{\%Arrange}

For the proportion of time that participants spent on the Arrange section of the test cases, our results indicate that their experience with Java has a statistically significant effect on it, while we do not find statistically significant results with regard to participants' experience in using tests and or their prior knowledge. The \textit{\%Arrange} variable is positively affected by the amount of experience in Java in years.

\subsubsection{\%Act}

For the proportion of time that participants spent on the Act section of the test cases, our results do not show any statistically significant effect of the independent variables used in this study on the dependent variable. 

\subsubsection{\%Assert}

For the proportion of time that participants spent on the Assert section of the test cases, our results indicate only one statistically significant effect, namely for the independent variable of experience with Java. The \textit{\%Assert} variable is negatively affected by the amount of experience in Java in years.

\subsubsection{Purpose Score (ITP)}

With regard to participants' ability to identify the purpose of the provided test suite, our results indicate only one statistically significant effect, namely for the independent variable of experience with using tests. The odds of identifying the testing purposes are positively affected by experience with using tests.

\subsubsection{Producing Basic Test Cases}

With regard to participants' ability to produce at least one additional test case that satisfies the criteria of being a basic test case, our results do not indicate any statistically significant relationship.

\subsubsection{Producing Domain Test Cases}

With regard to participants' ability to produce at least one additional test case that satisfies the criteria of being a domain-related test case, our results indicate two independent variables with statistically significant positive influence: the participants' experience with using tests and whether they have prior knowledge of the software. 

\subsubsection{Producing Error Test Cases}

With regard to participants' ability to produce at least one additional test case that satisfies the criteria of testing an error, our results indicate two factors with statistically significant positive influence: the participants' experience in Java and the total reading time on the particular test suite. The latter is too small to be considered though (an increase by 1.00001 times only).

\subsubsection{Producing Other Test Cases}

With regard to participants' ability to produce at least one additional test case that did not fit other categories, our results indicate one factor with a statistically significant effect: the participants' experience with using tests. The odds of producing other test cases are positively affected by experience with using tests.

\section{\label{cha:discussion}Discussion}

\begin{table*}
\centering
\caption{Dependent variables with their respective independent variables that have a statistically significant influence, if any. Positive influences are depicted using $\uparrow$ and negative influences are depicted using $\downarrow$. The table includes the parameters for each model.}
\label{tbl:significant_variables}
\begin{tabular}{llrlrrrrl}
\toprule
Dependent & Independent & p-value & Effect & $\chi^2(1)$ & Marginal & Conditional & McFadden     & Log Effect                             \\
Variable  & Variables   &         &        &             & $R^2$    & $R^2$       & pseudo-$R^2$ &                                        \\
\midrule
Reading Time       & Project Knowl.~($\downarrow$)          & 0.022   & $-1.815 \pm 1.284$ sec.     & 5.364       & 0.165          & 0.726             & --                    & $-0.259 \pm 0.109$  \\
\%Arrange          & Java ($\uparrow$)                      & 0.009   & $+0.867\% \pm 0.328\%$      & 6.73        & 0.049          & 0.586             & --                    & --                 \\
\%Act              & --                                     & --      & --                          & --          & 0.1            & 0.298             & --                    & --                 \\
\%Assert           & Java ($\downarrow$)                    & 0.020   & $-0.636\% \pm 0.276\%$      & 4.92        & 0.021          & 0.776             & --                    & --                 \\
Purpose            & Tests ($\uparrow$)                     & 0.008   & $+e^{0.272} = 1.313$ times  & --          & --             & --                & 0.216                 & $+0.272 \pm 0.103$ \\
Basic Test Cases   & --                                     & --      & --                          & --          & --             & --                & 0.052                 & --                 \\
Domain Test Cases  & Tests ($\uparrow$)                     & 0.002   & $+e^{0.339} = 1.404$ times  & --          & --             & --                & 0.157                 & $+0.339 \pm 0.108$ \\
                   & Project Knowl.~($\uparrow$)            & 0.001   & $+e^{1.975} = 7.207$ times  & --          & --             & --                & 0.157                 & $+1.975 \pm 0.619$ \\
Error Test Cases   & Java ($\uparrow$)                      & 0.023   & $+e^{0.345} = 1.411$ times  & --          & --             & --                & 0.279                 & $+0.345 \pm 0.152$ \\
Other Test Cases   & Tests ($\uparrow$)                     & 0.024   & $+e^{0.266} = 1.304$ times  & --          & --             & --                & 0.153                 & $+0.266 \pm 0.118$ \\
\bottomrule
\end{tabular}
\end{table*}

In this section, we revisit our research questions, formulate answers based on the obtained results, elaborate on how our findings compare against relevant literature, and provide concrete recommendations to software developments teams and researchers based on our work. A summary of the significant factors identified in this work can be found in Table~\ref{tbl:significant_variables}. 

\subsection{RQ1. What factors influence the time that developers spend reading test code?}

The only factor for which we found a statistically significant influence on reading time was whether participants had prior knowledge of the software project. Having prior knowledge of the software project reduces the time that developers spend reading test code. Contrary to similar studies focusing on aspects relevant to reading source code~\cite{crosby2002roles, Hofmeister2018, hofmeister2017comparing, Sharif:camelcase, kelly2015using, Busjahn:2015:EMC:2820282.2820320, Sharif:2012:ESR:2168556.2168642, Schankin:2018:DCI:3196321.3196332,peachock2017investigating}, we did not find differences in the reading time between experts and novices. While observing significant differences based on experience level is common in studies in this field, none of the three types of considered experience in this study show significant influence on the reading time of participants.

\implicationo{Having prior knowledge of the software under test significantly reduces the time developers have to spend to comprehend its test code.}

Based on the distribution of time that participants spent on each section of the AAA test structure, several observations can be made. In general, developers spent the least amount of their time on the Act section of test classes, while most of their time is spent on the Arrange section.

\implicationo{Developers spend the majority of their time on comprehending the inputs of a given test (the Arrange part), and little time on actually comprehending which method is under test (the Act part).}

We found only one statistically significant factor of influence on the time that developers spend reading the Arrange part of test code: their experience with the Java programming language. The time spent reading the Arrange part of a test case is positively affected by developers' amount of experience with Java. Our results do not indicate any factor with significant impact on the time spent by developers on reading the Act part of a test case. Similar to the Arrange part, we found only one statistically significant factor of influence on the time that developers spend reading the Assert part of test code: the amount of Java experience of participants, which negatively affects the time. We speculate that developers with more Java experience are more likely to care about the exact inputs being tested, thus spending more time on the Arrange section, and that they might be more familiar with the Assert syntax, possibly explaining the lower amount of time spent on test assertions.

\implicationo{Experience with Java reduces the amount of time spent on reading test assertions. }

\implicationo{Experience with Java increases the amount of time spent on reading the Arrange section of a test case. }

\subsection{RQ2. What factors influence the ability of developers to identify the purpose of a test suite? (ITP)}

The only factor for which we found a statistically significant influence on ITP is the experience of participants with using tests. Every additional year of experience with using tests has a positive effect on the odds of correctly identifying the purpose of a test suite.

\implicationo{Experience with tests significantly improves test code comprehension.}

\subsection{RQ3. What factors influence the ability of developers to produce additional test cases to extend a test suite? (PAC)}

Our research revealed no statistically significant influential factors on developers' ability to produce basic test cases (i.e., test cases only based on information in the provided test cases). 

\implicationo{We have no evidence to suggest that the ability of developers to produce basic test cases is correlated to any type of experience.}

Prior knowledge of the software project and the experience of participants with using tests have positive effects on the ability of participants to produce domain test cases (i.e., test cases extending the provided test suite with scenarios pertinent to the software project).

\implicationo{Prior knowledge of the software project is fundamental for domain-related test cases (even in domains with which most people are already somewhat familiar).}

Lastly, the influential factor for being able to produce error test cases (i.e., test cases validating invalid or erroneous scenarios) is the experience with the Java programming language, having a positive effect.

\implicationo{Experience with programming language significantly impacts the amount of error test cases a developer can think of.}

These findings are in line with existing literature stating the differences in program comprehension between developers with different levels of experience~\cite{crosby2002roles, Hofmeister2018, hofmeister2017comparing, Sharif:camelcase, kelly2015using, Busjahn:2015:EMC:2820282.2820320, Sharif:2012:ESR:2168556.2168642, Schankin:2018:DCI:3196321.3196332, peachock2017investigating}. However, our findings differ from existing literature by going beyond solely finding these differences based on experience level. Particularly, compared to existing studies, our work also investigates the impact of different types of experience (i.e., as a developer, with the Java programming language, and with using tests) and states the specific impact that independent variables have on different metrics of test code comprehension. All dependent variables regarding test code comprehension for which at least one significant relationship was found, are positively impacted by at least one type of experience. Our work allows for the observation of the differences in impact of different types of experience.

\implicationo{Different types of experience influence different activities related to test code comprehension.}

\subsection{Practical Recommendations}

In the following, we provide recommendations for software development teams and researchers based on our work:

\begin{itemize}

\item We know that, in practice, developers rarely know their entire source code bases (and even may forget what they have seen/known before~\cite{kruger2018you}). Nevertheless, we reinforce the recommendation for software development teams to invest in internal education, as our results suggest that source code familiarity is related to faster test code comprehension. We see future research on recommending developers where to spend their learning effort (e.g.,~\cite{fritz2010degree}). 

\item Developers are already aware of test code smells~\cite{meszaros2007xunit} and of how important test code quality is for software maintenance (e.g.,~\cite{Bavota:2015:TSR:2790618.2790636}). Our work adds to existing test smell catalogues as it shows how important the Arrange part of the test is for the comprehension of the test itself. Thus, we recommend developers to make sure that the Arrange parts of their tests are clear, concise and self-explanatory. Future research should focus on how DSLs such as AssertJ\footnote{\url{http://joel-costigliola.github.io/assertj/}} and the Test Data Builder pattern\footnote{\url{http://www.natpryce.com/articles/000714.html}} improve readability. 

\item Seniority in the project plays a role in developers' capacity of comprehending test code. Our results show that newcomers have more trouble in developing domain-related test cases. The experience of developers is known to play a role in other software maintenance activities, such as code reviews~\cite{jiang2013will}. We see a great importance for approaches such as the one proposed by Pham et al.~\cite{pham2015automatically} where newcomers receive examples of test code written by senior developers as a way to learn more about the project. However, given the current state of the practice, we suggest experienced developers to review the test code from novices, as odds are that they will not test domain cases (and error cases, in case of language novices). 

\item Our study introduced two new metrics for test code comprehension: the capacity of developers to produce additional test cases (PAC) and the capacity of developers to identify the overall purpose of a test suite (ITP). As far as we know, we are the first to propose these metrics. We strongly believe these variables are important when it comes to test code comprehension. Future research needs to 1) investigate how to better measure these variables, and to 2) use these metrics in future test code comprehension studies. 

\end{itemize}

\section{\label{sec:validity}Threats to Validity}

In this section, we discuss the threats that could affect the validity of our results and conclusions.

\subsection{Internal Validity}

The participants were given a maximum amount of time to read the provided test code, contrary to scenarios in realistic environments, and there is a possibility that participants felt rushed by the timer, resulting in the answers to the open questions to be of lower quality. On the other hand, inspection of the respective distribution in Table \ref{tbl:dependent_numerical} shows that it is not skewed towards the upper time limit (600 seconds). None of the participants spent the full amount of time in any of the trials and during 97\% of the recorded trials, less than 400 seconds were spent on reading the provided test code. We conclude that the time limit of ten minutes had no important influence on the quality of our results.

Participants did not have access to the production code when reading the test code to enable us to clearly identify whether the answers provided by developers were given due to knowledge they extracted from the production or from the test code. In future work, we will study how test code comprehension is affected by the presence of production code.

\subsection{Construct Validity}

In this research, we have associated and measured test code comprehension with three different factors, namely the total time that participants spent on reading the test suite (RT), their ability to identify the testing purpose of a test suite (ITP), and their ability to produce additional cases to extend the test suite (PAC). While the factor of time is a representative partial metric of program comprehension, as used by previous studies~\cite{Hofmeister2018, hofmeister2017comparing, Sharif:camelcase, Jbara:2014:ECR:2597008.2597140, Melo:2017:VTE:3101414.3101419, Sharif:2012:ESR:2168556.2168642, Schankin:2018:DCI:3196321.3196332}, no other studies have investigated factors similar to ITP and PAC. We manually evaluated the participants' answers to the ITP and PAC questions, and thus, our results depend on the quality of our manual work. Our online appendix~\cite{appendix} contains the raw data for inspection.

We did not use eye tracking as a tool to track where developers are looking in a test class, but rather created our own tracking software with a limited 5-line window for this purpose, similar to studies by Schankin et al.~\cite{Schankin:2018:DCI:3196321.3196332} and Hofmeister et al.~\cite{Hofmeister2018, hofmeister2017comparing}. The main trade-off that contributed to this decision was between quality and quantity of the data to be gathered. While performing an eye tracking study yields major benefits in terms of the quality, it comes with major drawbacks on how and how much data can be acquired. While the fine-grained data (which part of the tests developers were looking at) is only used as one variable in the logistic regression, many of the studied variables (time, ITP, and PAC) do not require it. Participants had to scroll back and forth to see a complete test case, which may have affected their willingness to look at different sections of the test code.

\subsection{External Validity}

All participants were volunteers, which possibly has influence on the results due to volunteers generally being more motivated. The participants with knowledge of the code base were students, adding potential bias to the comparison between participants with project knowledge and those without. It is also possible that the student participants were influenced by having participated in our courses before. A scenario where developers develop tests for a project which they are not familiar with may be artificial, although casual contributors have been found to add test cases to open source projects, for example~\cite{Egua}. Whether participants had prior knowledge of the software project is a factor that has rarely been investigated in research towards program comprehension, even though it can have considerable influence (as shown by our results). We believe our results are a first step in raising the importance of this feature for future studies. Participants had to first read the code and answer questions subsequently, which does not represent a realistic scenario and might affect generalizability. Finally, we used test code snippets from \jpacman, a small project with a well-designed test suite in terms of code quality. Although developers face more complicated test code in real life, providing them with a well-designed suite gives us a baseline for comparison in future studies. More realistic settings, such as test suites with smells~\cite{DeursenMBK01}, test suites of applications with complex problems, and test code at different test levels (e.g., integration testing), are part of our future work.

\section{\label{cha:conc}Conclusion}

We present an empirical study to investigate factors which influence test code comprehension of software developers. To measure test code comprehension, we decompose it into three metrics: (i) how much time developers spend on reading a test suite, (ii) whether developers are able to correctly identify the testing purpose of a test suite, and (iii) their ability to provide additional test cases to an existing test suite.

The main findings based on our results are that (i) having prior knowledge of the software project decreases the amount of time developers spend on reading the provided test suite, (ii) experience with the Java programming language affects the proportions of time spent on the Arrange and Assert sections of tests, (iii) experience with the Java programming language and having prior knowledge of the software project increases the likelihood of producing certain categories of additional test cases, and (iv) the most positive influential factor towards understanding and extending a test suite is experience with using tests.

To the best of our knowledge, this is the first study to investigate the process of program comprehension with a focus on testing, and it provides a basis for future work towards understanding of test code comprehension.

\section*{Acknowledgements}

This work has in part been supported by the Australian Research Council's Discovery Early Career Researcher Award (DECRA) funding scheme (DE180100153).


\end{sloppy}
\end{document}